\documentclass[epj,referee]{svjour}
\usepackage{bm}
\usepackage{graphicx}

\begin{document}

\title{Low temperature spin glass fluctuations: \\ 
       expanding around a spherical approximation}

\author{A. Crisanti\inst{1} \and C. De Dominicis\inst{2} \and T. Sarlat\inst{2}}

\institute{ \inst{1} Dipartimento di Fisica, Universit\`a di Roma 
                     {\em La Sapienza} and  SMC, 
                     P.le Aldo Moro 2, I-00185 Roma, Italy.\\ 
            \inst{2} {\em Institut de Physique Th\'eorique}, CEA -
                      Saclay - Orme des Merisiers, 91191 Gif sur Yvette, 
                     France.\\ 
}

\date{Received: V 7.5.4 2009/11/03 15:56:09 AC / Revised version: date} 

\abstract{
The spin glass behavior near zero temperature is a complicated matter.
To get an easier access to the 
spin glass order parameter $Q(x)$ and, at the same time, keep track
of $Q_{ab}$, its matrix aspect, and hence of the Hessian controlling stability,
we investigate an expansion of the replicated free energy functional around its
``spherical'' approximation. This expansion is obtained by introducing a
constraint-field and a (double) Legendre Transform expressed in terms of
spin correlators and constraint-field correlators. The spherical approximation 
has the spin fluctuations treated with a global constraint and the 
expansion of the Legendre Transformed functional brings them closer and closer 
to the Ising local constraint. In this paper we examine the first 
contribution of the systematic corrections to the 
spherical starting point.
}


\maketitle

\section{Introduction}
\label{sec:s0}

The infinite range Sherrington-Kirkpatrick (SK) model is defined by
\cite{SheKir78}:
\begin{eqnarray}
{\cal H}_{\rm SK}=-\frac{1}{2} \sum_{ij}^{1,N} J_{ij} \sigma_i \sigma_j
\end{eqnarray}
where  the $\sigma_i$'s are $\pm 1$ Ising spins and 
the couplings $J_{ij}$ independent Gaussian random variables with null 
mean and variance equal to $1/N$. This model is solved with the 
replica trick, that is all the thermodynamical information is encoded in
the $n\to 0$ limit of the disorder-averaged replicated partition
function:
\begin{eqnarray}
\overline{Z^n}&\equiv& \overline{{\rm Tr}_{\sigma_{ia}} 
\exp\left(-\beta \sum_{a=1}^n 
{\cal H}_{\rm SK}({\bm \sigma}_{a})\right)}
\end{eqnarray} 
where $\beta = 1/T$ is the inverse of temperature and, as usual, 
$\overline{(\cdots)}$ the average over the quenched disorder. 
Performing this average, and introducing the auxiliary symmetric 
replica matrix
$Q_{ab}\equiv \frac{1}{N}\sum_{i=1}^N\, \sigma_{ia}\,\sigma_{ib}$ with 
$a\not=b$, leads to \cite{SheKir78}:
\begin{eqnarray}
\overline{Z^n}= \int \prod_{ab} \sqrt{\frac{N\beta^2}{2\pi}}\,dQ_{ab}
             \, e^{N {\cal L}[Q]}
\label{eq:lag}
\end{eqnarray}
with the effective Lagrangian (density) \cite{note:omegaq}:
\begin{eqnarray}
{\cal L}[Q]&=& -\frac{\beta^2}{4} \sum_{ab} Q_{ab}^2 
                       + \Omega[Q]
\label{eq:l} \\
\Omega[Q] &=&\ln {\rm Tr}_{\sigma_a} \exp\bigg(\frac{\beta^2}{2}\sum_{ab} 
             Q_{ab}\, \sigma_a\sigma_b\bigg)\ .
\label{eq:ome}
\end{eqnarray}
The normalization factor in eq. (\ref{eq:lag}) gives a sub-leading 
contribution for $N\to\infty$ and is omitted in the following.

In the thermodynamical limit, $N\to \infty $, the value of the
integral in (\ref{eq:lag}) is given by the saddle-point value
$ -n\beta f_{\rm SK} = {\cal L}[Q]$, where $Q_{ab}$ is evaluated from 
the stationarity condition:
\begin{eqnarray}
\frac{\delta}{\delta Q_{ab}} {\cal L}[Q] = 0, \qquad a\not= b
\label{eq:sta}
\end{eqnarray}
that is,
\begin{equation}
Q_{ab} = \langle \sigma_a \sigma_b\rangle\ .
\end{equation}
The average is taken with respect to the weight
\begin{equation}
\zeta({\bm\sigma}) = 
      e^{\frac{\beta^2}{2}\sum_{ab} Q_{ab}\, \sigma_a\sigma_b }
      \Big/
      {\rm Tr}_{\bm\sigma}
      e^{\frac{\beta^2}{2}\sum_{ab} Q_{ab}\, \sigma_a\sigma_b}\ .
\label{eq:zet}
\end{equation}
If instead, one had in mind to write the effective Lagrangian governing a 
short-range ({\it e.g.} nearest-neighbor) system, one would consider:
\begin{eqnarray}
e^{{\cal W}[H] }&=& \int \prod_{i,ab} dQ_{ab}(i) 
\exp\bigg(-\frac{\beta^2}{4}\sum_p \sum_{ab} (p^2+1)\, Q_{ab}^2(p) 
\nonumber \\
&\phantom{=}&
 + \sum_{i}\Omega[Q(i)] + \frac{1}{2}  \sum_{i,ab}H_{ab}(i)\,Q_{ab}(i)\bigg)\ .
\label{eq:ltr}
\end{eqnarray}
Here $Q_{ab}(p)$ is the (space) Fourier Transform of the site-dependent
replica overlap matrix $Q_{ab}(i)$, with $i$ the site-index, 
$H_{ab}(i)$ is an external (unphysical) field that couples 
to $Q_{ab}(i)$, and $\Omega[Q(i)]$ is given for each site $i$ 
by (\ref{eq:ome}).
Properties of ${\cal W}[H]$, and of its Legendre Transform 
$\Gamma[Q]$, defined {\it via}:
\begin{eqnarray}
{\cal W}[H] + \Gamma[Q] = \frac{1}{2} 
     \sum_{i,ab}H_{ab}(i)\,Q_{ab}(i) 
\end{eqnarray}
and, 
\begin{eqnarray}
\frac{\delta}{\delta Q_{ab}(i) } \Gamma[Q] = H_{ab}(i)
\end{eqnarray}
are obtained by writing:
\begin{eqnarray}
Q_{ab}(i)=Q_{ab}+\delta Q_{ab}(i)
\end{eqnarray} 
with $Q_{ab}$ given as the (site-independent) saddle-point va\-lue of 
${\cal L}$, eq. (\ref{eq:sta}). An expansion of 
(\ref{eq:ltr}) in powers of $\delta Q_{ab}(i)$ shows that no linear terms 
({\it i.e.} tadpoles) remains by vir\-tue of (\ref{eq:sta}). 
The quadratic term is given by:
\begin{eqnarray}
\frac{1}{8T^2} \sum_p\sum_{abcd} \delta Q_{ab}(p)\, 
     {\cal M}^{ab,cd}(p) \, \delta Q_{cd}(p) 
\end{eqnarray}
where ${\cal M}$ is the Hessian matrix whose eigenvalues are the 
so-called bare masses of the $\delta Q$ correlation functions 
({\it i.e.} the inverse propagator). With these notations, one has:
\begin{eqnarray}
{\cal M}^{ab,cd}(p)=(p^2+1)\,\delta^{kr}_{ab,cd} 
   - T^2\,\frac{\delta^2}{\delta Q_{ab}\delta Q_{cd} } \Omega[Q]\ .
\end{eqnarray}
The higher order terms are in turn the couplings (cubic, quartic,...) of the 
fields $\delta Q$.

Whatever the adopted viewpoint, one cannot avoid the difficult construction of 
the 
$\Omega$ functional as given in (\ref{eq:ome}). 
Parisi and others \cite{Parisi79,Parisi80,Duplantier81} have shown 
how to obtain solutions with $R$ steps of Replica Symmetry Breaking (RSB) and 
in particular with $R\to\infty$, and how to construct equations satisfied by 
$Q(x)$, the continuous limit of the order parameter $Q_{ab}$
for $R\to \infty$ \cite{Parisi83}. 
These equations can be solved in the full low temperature
phase \cite{CriRiz02}.
Clearly, however, for very low temperatures or for null temperature, the 
problem is delicate, as one can experience when expanding $\zeta({\bm \sigma})$
in powers of $1/T^2$ [see eq. (\ref{eq:zet})]. 
The properties of the Parisi solution $Q(x)$ at very low temperatures have
attracted some work in last years. The analysis has been carried on either
directly on the continuous limit, i.e. taking first $R\to\infty$ an then 
$T\ll 1$ \cite{CriRizTem03,Pankov06}, or on the $T\to 0$ limit of large
$R$-step replica symmetry breaking solutions 
\cite{OppShe05,OppSchShe07,SchOpp08,Schmidt08}.
In both approaches, however, properties of the Hessian remain elusive since 
once one is working with $Q(x)$ the matrix structure of the order parameter 
$Q_{ab}$ is lost. On the other hand taking first $T\to 0$ and then $R\gg 1$ one 
encounters problems following the delicate commutativity of the two limits, 
extensively discussed in Ref. \cite{OppSch08}.
These two faced difficulties on the road to obtaining the Hessian 
(and eigenvalues) have motivated this work.
The starting point is the observation that, if the strict ``local'' 
constraint $\sigma_a^2=1$ is replaced, as a first step, by the global 
constraint 
$\sum_a\,\sigma_a^2 = n$ (a kind of {\sl spherical approximation} in
replica space) then, the temperature behavior gets simpler while keeping the 
matrix structure of the problem. We show below how to systematically
return towards the local constraint introducing higher and higher 
correlations 
generated under a double Legendre Transform on the $\Omega$ functional.
In the resulting approximated functional for $\Omega$, the null temperature 
limit becomes non-singular, and, at the same time, one keeps the matrix
structure of $Q_{ab}$, leaving access to the Hessian.

The outline of the paper is as follows: in Section \ref{sec:gen} we describe
the general formalism based on the double Legendre Transform and the stability
of the saddle-point. In Sections \ref{sec:zero} and \ref{sec:twor} we 
discuss respectively the (trivial) zeroth order approximation and the first
non-trivial approximation, the two-replica approximation, of the functional
$\Omega$. 
To make the paper also accessible to uninitiated readers,
we have deferred to appendices, besides a couple of detailed calculations,
a reminder concerning some of the technical tools  used in the text
(Replica Fourier Transform and double Legendre Transform).

\section{General formalism}
\label{sec:gen}
\subsection{Equation of motion: }
For spherical models, the spin variables $\sigma_a$ are continuous:
the trace over the spins becomes an $n$-dimensional
Gaussian integral. Here, the Ising spins are discrete variables 
$\sigma_a=\pm 1$. To overcome the discrete 
nature of the spin, one introduces a constraint-field:
\begin{eqnarray}
\label{eq:trace}
{\rm Tr}_{\bm\sigma}(\cdots) &=& \prod_{a=1}^{n}\sum_{\sigma_a=\pm 1} (\cdots)
\nonumber\\
                     &=& \prod_{a=1}^{n}\int_{-\infty}^{+\infty}
     d\sigma_a\,\left[\delta(\sigma_a+1) + \delta(\sigma_a-1)\right] (\cdots)
\nonumber\\
     &=& \int_{-\infty}^{+\infty}\prod_{a=1}^{n} d\sigma_a
       \int_{-i\infty}^{+i\infty}\prod_{a=1}^{n} \frac{d\lambda_a}{2\pi i}
\nonumber\\
&\phantom{=}& \times
       \exp\left[-\frac{1}{2}\sum_a\,\lambda_a\left(\sigma_a^2-1\right)\right]
       (\cdots)\ .
\end{eqnarray}
The auxiliary variables $\lambda_a$ control the fluctuations of the continuous
spin variables $\sigma_a$ around the Ising values $\pm 1$ and thus we expect 
that they diverge as $T\to 0$.

By inserting (\ref{eq:trace}) into the expression (\ref{eq:ome}), 
the functional ${\cal L}[Q]$ can be written as:
\begin{equation}
{\cal L}[Q] = 
                  - \frac{\beta^2}{4}\sum_{ab}^{1,n}\,Q_{ab}^2
                  + \Omega_{0^+}[Q]
                  - \frac{n \beta^2}{4}
\end{equation}
where
\begin{eqnarray}
\Omega_{0^+}[Q] &=& \lim_{\epsilon\to 0^+} \Omega_{\epsilon}[Q] 
\nonumber\\
            &=& \lim_{\epsilon\to 0^+} \ln\int_{-\infty}^{+\infty}d{\bm\sigma}
               \int_{-i\infty}^{+i\infty}\frac{d{\bm\lambda}}{(2\pi i)^n}\,
   e^{ -S[{\bm\sigma},{\bm\lambda}]}
\label{eq:zetae}
\end{eqnarray}
and $S[{\bm\sigma},{\bm\lambda}]$ is given by,
\begin{eqnarray}
S[{\bm\sigma},{\bm\lambda}] &=& 
 -\frac{\beta^2}{2}\sum_{ab}Q_{ab}\,\sigma_a\sigma_b
             - \frac{\epsilon}{2}\sum_a\lambda_a^2 
\nonumber\\
&\phantom{=}&
             + \frac{1}{2}\sum_a\,\lambda_a\sigma_a^2 
             -\frac{1}{2}\sum_a\lambda_a\ .
\label{eq:esse}
\end{eqnarray}
The constant term $-n\beta^2/4$ follows from the (definition of the) 
diagonal terms $Q_{aa}=1$
of the overlap matrix. 
We recall that when introduced, the overlap matrix 
is defined only for $a\not=b$, the reason being that for Ising 
spins $\sigma^2=1$. The diagonal terms $Q_{aa}$ are then usually (tacitly) 
set to zero, as in the previous Sections. 
This choice may not be the most suitable for spherical models 
where sums over spin values are replaced by Gaussian integrals. 
We thus (re)define the overlap matrix with the diagonal terms equal to $1$.
The constant term comes out then for consistency with eqs. (\ref{eq:l}) and 
(\ref{eq:ome}).

The functional $\Omega_{\epsilon}[Q]$ can be expressed in terms of the
field expectation values and two-point correlation functions
by introducing linear and bilinear auxiliary fields 
coupled with ${\bm \sigma}$ and ${\bm \lambda}$, and
performing a double Legendre Transform with respect to these fields 
\cite{DeDomMar64,CorJacTom74,Haymaker91}.
See also Appendix \ref{app:DLT}.
In absence of an external field coupled to $\sigma_a$  the average 
$\langle{\sigma_a}\rangle$ vanishes. Moreover, one can check that 
$\langle\sigma_a\lambda_b\rangle = 0$ \cite{note:slcorr}. Therefore
the relevant field averages and correlation functions are \cite{note:lambda}
\begin{equation}
\langle{\lambda_a}\rangle = \lambda
\end{equation}
and
\begin{equation}
\langle{\sigma_a\sigma_b}\rangle = G_{ab}, \quad
\langle{\lambda_a\lambda_b}\rangle = \Lambda_{ab}\ .
\end{equation}
All averages are evaluated with the weight $\exp(-S[{\bm\sigma},{\bm\lambda}])$.

After the double Legendre Transform has been taken, and the auxiliary
linear and bilinear fields set to zero, the functional 
$\Omega_{\epsilon}[Q]$ comes out to be \cite{CorJacTom74}:
\begin{eqnarray}
\Omega_{\epsilon}[Q] &=& -S[{\bm 0},{\bm \lambda}] 
        - \frac{1}{2}{\rm Tr}\ln G^{-1}
        - \frac{1}{2}{\rm Tr}\ln \Lambda^{-1}
\label{eq:omedlt}\\
&\phantom{=}&
        - \frac{1}{2}{\rm Tr}D^{-1}G 
        - \frac{1}{2}{\rm Tr}\Delta^{-1}\Lambda
        - {\cal K}_2[G,\Lambda] + n 
\nonumber
\end{eqnarray}
where,
\begin{equation}
\label{eq:freep}
(D^{-1})_{ab} = \lambda\delta_{ab} - \beta^{2}\,Q_{ab}, \quad
(\Delta^{-1})_{ab} = -\epsilon\,\delta_{ab}\ .
\end{equation}
Here, ${\cal K}_2[G,\Lambda]$ is given by the sum of all the two-particle 
irreducible (2-PI) vacuum graphs in a theory with vertices 
\begin{equation}
  -\frac{1}{2}\lambda_a\sigma_a^2  =\ %
\raisebox{-18pt}{\includegraphics{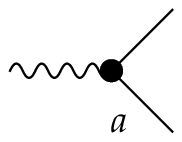}}
\end{equation}
and propagators 
\begin{equation}
G_{ab}  =\ %
\raisebox{-15pt}{\includegraphics{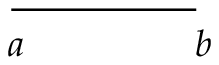}}\ ,
\qquad
\Lambda_{ab}  =\ %
\raisebox{-15pt}{\includegraphics{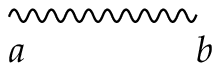}}\ .
\end{equation}
The first terms of the diagrammatic expansion of ${\cal K}_2[G,\Lambda]$ are:
\begin{eqnarray}
{\cal K}_2[G,\Gamma]  &=&\quad %
\raisebox{-20pt}{\includegraphics{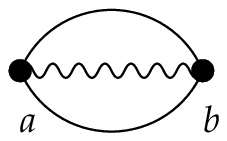}} 
\nonumber\\
&\phantom{=}& + %
\raisebox{-39pt}{\includegraphics{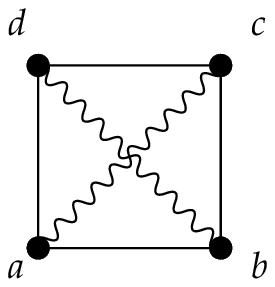}} + \cdots \ .
\end{eqnarray}
By construction,
when the auxiliary fields are set to zero the double Legendre Transform 
is stationary with respect to the variations of
$\lambda$, $G$ and $\Lambda$. This leads respectively to:
\begin{eqnarray}
G_{aa} &=& 1
\label{eq:gaa}
\\
(G^{-1})_{ab} &=&  (D^{-1})_{ab} +
     2\frac{\partial}{\partial G_{ab}}{\cal K}_2[G,\Lambda]
\label{eq:dg}
\\
(\Lambda^{-1})_{ab} &=&
     2\frac{\partial}{\partial \Lambda_{ab}}{\cal K}_2[G,\Lambda]\ .
\label{eq:dgam}
\end{eqnarray}
Equations (\ref{eq:dg}) and (\ref{eq:dgam}) are known as the Dyson equation 
for the 
two-point functions $G$ and $\Lambda$ with self-energies:
\begin{eqnarray}
\Sigma_{ab}^{\sigma\sigma} &=& 
          -2\frac{\partial}{\partial G_{ab}}{\cal K}_2[G,\Lambda] \\ 
\Sigma_{ab}^{\lambda\lambda} &=& 
          -2\frac{\partial}{\partial \Lambda_{ab}}{\cal K}_2[G,\Lambda]\ .
\end{eqnarray}
These Dyson equations result from the Legendre Transform, and give {\it exact} 
relations for $G$ and $\Lambda$. By solving them, the propagators $G_{ab}$ and 
$\Lambda_{ab}$ become functionals of $Q_{ab}$ that, substituted back into
(\ref{eq:omedlt}), give
$\Omega[Q]$ as functional of $Q_{ab}$ alone.
By using now the saddle-point equation (\ref{eq:sta}) one
can write the self-consistent equation for $Q_{ab}$, that reads
\begin{equation}
Q_{ab} = G_{ab}, \quad a\not=b\ .
\label{eq:gab}
\end{equation}
With the choice $Q_{aa}=1$ eqs. (\ref{eq:gaa}) and (\ref{eq:gab}) 
can be merged into the single equation of motion
\begin{equation}
\label{eq:gabg}
  Q_{ab} = G_{ab}\quad \forall a,b\ .
\end{equation}
We stress that, while in the following we use the stationarity conditions
in this form, one should keep in mind that the diagonal and off-diagonal 
terms of these equations follow from two distinct conditions. The diagonal 
terms follow from the double Legendre Transform, 
and our choice $Q_{aa}=1$. This is an exact relation.
The off-diagonal terms follow from the saddle point eq. (\ref{eq:sta}) and 
hence eq. (\ref{eq:gab}) is valid only in the thermodynamic limit.

\subsection{Stability}
To decide whether the solutions of the equation of motion are stable or not, 
one has to expand the effective Lagrangian to second order in $\delta Q$:
\begin{eqnarray}
{\cal L}\{Q_{ab}+\delta Q_{ab}(p)\} &=& {\cal L}^{(0)}\{Q_{ab}\} 
\nonumber\\
&\phantom{=}& \hskip-2cm
+ \frac{\beta^2}{2} \sum_{ab} \delta Q_{ab}(0) (G_{ab}-Q_{ab})
\nonumber\\
&\phantom{=}& \hskip-2cm
-\frac{\beta^2}{8} \sum_p\sum_{abcd} \delta Q_{ab}(p) {\cal M}^{ab,cd}(p) 
\delta Q_{cd}(p) 
\end{eqnarray}
with the Hessian matrix:
\begin{eqnarray}
{\cal M}^{ab,cd}(p)=(p^2+1)\,\delta^{\rm Kr}_{ab,cd} - 
\frac{\delta G_{ab}}{\delta Q_{cd} }\ .
\end{eqnarray}
Let us introduce:
\begin{eqnarray}
{\cal X}^{ab,cd}&\equiv&\frac{\delta G_{ab}}{\delta Q_{cd} } \\
{\cal Y}^{ab,cd}&\equiv&\frac{\delta \Lambda_{ab}}{\delta Q_{cd} } 
\end{eqnarray}
which through the equations of motion (\ref{eq:dg}) and (\ref{eq:dgam})
satisfy:\footnote{Please note that in eqs. 
(\protect{\ref{eq:xsta}})-(\protect{\ref{eq:eq39}}) we used
$G^{-1}_{ab}$ for $(G^{-1})_{ab}$ to bypass cumbersome writing.}
\begin{eqnarray}
&-&\sum_{ef}\bigg[G^{-1}_{ae}G^{-1}_{bf}+G^{-1}_{af}G^{-1}_{be}
+\frac{\delta^2}{\delta  G_{ab}\delta  G_{ef}} {\cal K}_2\bigg]{\cal X}^{ef,cd}
\nonumber\\ 
&\phantom{-}& \hskip0.5cm 
- \sum_{ef}
 \left(\frac{\delta^2}{\delta  G_{ab}\delta  \Lambda_{ef}} {\cal K}_2\right)\,  
{\cal Y}^{ef,cd}=-\frac{\delta^{\rm Kr}_{ab,cd}}{T^2}
\label{eq:xsta}
\\
&-&\sum_{ef}\bigg[\Lambda^{-1}_{ae}\Lambda^{-1}_{bf} +
\Lambda^{-1}_{af}\Lambda^{-1}_{be} + 
\frac{\delta^2}
{\delta  \Lambda_{ab}\delta  \Lambda_{ef}} {\cal K}_2\bigg]{\cal Y}^{ef,cd}
\nonumber\\
&\phantom{-}& \hskip0.5cm 
-\sum_{ef}
 \left(\frac{\delta^2}{\delta  \Lambda_{ab}\delta G_{ef}} {\cal K}_2\right)\,  
{\cal X}^{ef,cd}=0 \ .
\label{eq:ysta}
\end{eqnarray}
Solving (\ref{eq:ysta}) for ${\cal Y}$, and inserting it into (\ref{eq:xsta}), 
one has:
\begin{eqnarray}
&-&\sum_{ef}\bigg[G^{-1}_{ae}G^{-1}_{bf}+G^{-1}_{af}G^{-1}_{be}
 + \frac{\delta^2}{\delta  G_{ab}\delta  G_{ef}} {\cal K}_2
\nonumber\\
&\phantom{-}& 
-\sum_{gh}\frac{\delta^2}{\delta  G_{ab}\delta  \Lambda_{gh}} {\cal K}_2 
 \sum_{ij} {\cal D}^{-1}_{gh,ij} \frac{\delta^2}
      {\delta \Lambda_{ij}\delta  G_{ef}} {\cal K}_2 \bigg]{\cal X}^{ef,cd}
\nonumber\\
&\phantom{-}& 
=-\frac{\delta^{\rm Kr}_{ab,cd}}{T^2}\ .
\label{eq:eq39}
\end{eqnarray}
The matrix ${\cal D}$ is the one multiplying ${\cal Y}$ in (\ref{eq:ysta}). 
At this point, we can now use the equality $G_{ab}=Q_{ab}$, resulting from 
the stationarity condition.

To solve the self-consistent equations of motion, 
we have to specify the form of the 
matrix $Q_{ab}$ that takes into account possible breaking of the permutation 
symmetry of replica pairs. A standard parameterization has been introduced 
by Parisi \cite{Parisi79,Parisi80}.
The overlap matrix $Q_{ab}$ for $R$ allowed breaking in the replica 
permutation symmetry consists in dividing $Q_{ab}$ into successive boxes 
of decreasing size $p_r$, with $p_0 = n$ and $p_{R+1}=1$, along the diagonal 
and assigning the elements $Q_{ab}$ of the overlap matrix so that 
\begin{equation}
Q_{ab} \equiv Q_{a\cap b=r} \equiv Q_r, \qquad r = 0,\cdots, R+1
\end{equation}
with $1 = Q_{R+1} \geq Q_R \geq\cdots Q_1 \geq Q_0$. 
The notation  $a\cap b=r$ means that $a$ and $b$ belong to the 
same box of size $p_r$ but to two {\it distinct} boxes of size $p_{r+1} < p_r$.
The element $Q_0$ must vanish in 
absence of external fields that break the up/down symmetry \cite{note:q0}. 
In the following, if not explicitly stated, this will be always assumed.

In this formalism the Replicon component of the Hessian 
(that usually harbors the 
most dangerous, {\it i.e.}, the lowest eigenvalues), becomes
\cite{TemDeDomKon94,DeDomCarTem97}
\begin{eqnarray}
{\cal M}_{Rep}^{ab,cd}={\cal M}_{uv}^{rr} 
\end{eqnarray}
where $a\cap b\equiv c\cap d =r$ is the common overlap while
$u=\max\{a\cap c,a\cap d\}$ and $v=\max\{b\cap c,b\cap d\}$ the cross-overlaps.
Replicon matrices enjoy the property of being 
diagonalized under a double Replica Fourier Transform (RTF), 
and matrix products of 
such matrices becoming ordinary products of RFT matrices (like in ordinary 
Fourier Transform), we thus get \cite{DeDomCarTem97}:
\begin{eqnarray}
\label{eq:rep1}
{\cal M}_{\hat{k}\hat{l}}^{rr}&=&(p^2+1)- {\cal X}_{\hat{k}\hat{l}}^{rr}\\
\label{eq:rep2}
\frac{1}{{\cal X}_{\hat{k}\hat{l}}^{rr}}&=&
 T^2\bigg[\frac{1}{\widehat{G}_k\widehat{G}_l}
 + {\cal K}_{\hat{k}\hat{l}}^{G_rG_r}
 -{\cal K}_{\hat{k}\hat{l}}^{G_r\Lambda_r}
\frac{1}
{{\cal D}_{\hat{k}\hat{l}}^{rr}}{\cal K}_{\hat{k}\hat{l}}^{\Lambda_rG_r}\bigg]\\
\label{eq:rep3}
{\cal D}_{\hat{k}\hat{l}}^{rr}&=&\frac{1}{\widehat{\Lambda}_k\widehat{\Lambda}_l}
+{\cal K}_{\hat{k}\hat{l}}^{\Lambda_r\Lambda_r}
\end{eqnarray}
where $r=0,\ldots,R$ and $k,l= r+1,\ldots, R+1$.
By $ {\cal K}_{\hat{k}\hat{l}}^{G_rG_r}$ we mean the double RFT on the Replicon 
component of $\frac{\delta^2}{\delta  G_{ab}\delta  G_{cd}} {\cal K}_2$
with respect the lower indices (cross overlaps)
$u$ and $v$. The ``hat'' denotes the
RFT (see Appendix \ref{app:RFT}).

\section{Zero order approximation}
\label{sec:zero}
The zeroth order approximation is obtained by neglecting the 
2-PI contributions of ${\cal K}_2[G,\Lambda]$. 
Combining eqs. (\ref{eq:freep}), (\ref{eq:dg}) and (\ref{eq:gabg}) one is led to
\begin{equation}
\label{eq:eqRS}
(Q^{-1})_{ab} = \lambda \delta_{ab} - \beta^2\,Q_{ab}.
\end{equation}
To solve this equation the parameterization of the $Q_{ab}$ matrix is needed.
By using Parisi's parameterization 
one gets for the off-diagonal terms 
\begin{equation}
\label{eq:RS}
\left(Q^{-1}\right)_r = - \beta^2\,Q_r, \qquad r = 0,\ldots, R\ .
\end{equation}
As shown in Appendix \ref{app:proof}, for any $R$, 
this equation only admits the solution \cite{KosThoJon76}
\begin{equation}
\label{eq:RSsol}
Q_r = Q = 1 - T, \qquad \forall r
\end{equation}
i.e., the Replica Symmetric (RS) solution.
The value of $\lambda$ is set by the diagonal term $a=b$ of 
eq. (\ref{eq:eqRS}) and reads
\begin{equation}
\lambda = \frac{2}{T}\ .
\end{equation}
The stability of the solution is ruled by the Replicon eigenvalue
[see eqs. (\ref{eq:rep1})-(\ref{eq:rep3})],
\begin{equation}
{\cal M}^{00}_{\hat{1}\hat{1}}(p=0) = 1 - \beta^2(1-Q)^2 = 0,
\end{equation}
where $r=0$ and the (Fourier transformed) cross-overlap $k=l=1$.
The solution is then marginally stable, as known for the $2$-spin 
spherical model.

Let us note a technical point. In this approximation
$\Lambda_{ab} = -\epsilon\delta_{ab}$, so that the term 
${\rm Tr}\ln\Lambda$ is singular as $\epsilon\to 0^+$.
The functional ${\cal L}[Q]$ is nevertheless regular in this limit.
The reason is that neglecting the ${\cal K}_2[G,\Lambda]$ contribution 
is equivalent to remove the Ising-like constraint on the spin variables. This 
is also equivalent to discard the last two terms of
$S[{\bm\sigma},{\bm\lambda}]$ in eq. (\ref{eq:esse}). A
proper normalization is now needed to regularize the $\epsilon\to 0^+$
limit of $\Omega_{\epsilon}[Q]$. This introduces an additional constant term
in ${\cal L}[Q]$ that exactly balances the singularity coming 
from ${\rm Tr}\ln\Lambda$.

\section{Two-replica approximation}
\label{sec:twor}
The two-replica approximation consists in taking only the first
2-PI diagram in the diagrammatic expansion of ${\cal K}_2[G,\Lambda]$, namely,
\begin{eqnarray}
\label{eq:tworepl}
{\cal K}_2[G,\Lambda]  &=& 
\raisebox{-20pt}{\includegraphics{tworepl}}
\nonumber\\
&=&  \frac{1}{2}\left(-\frac{1}{2}\right)^2\,2
  \sum_{ab}\,G_{ab}^2\Lambda_{ab}\ .
\end{eqnarray}
The functional ${\cal L}[Q]$ then reads
\begin{eqnarray}
{\cal L}[Q] &=&   \frac{\beta^2}{4}\sum_{ab}\,Q_{ab}^2 
                + \frac{1}{2}{\rm Tr}\ln G
                + \frac{1}{2}{\rm Tr}\ln \Lambda
\nonumber\\
&\phantom{=}&
               \phantom{====} - \frac{1}{4}\sum_{ab}\,G_{ab}^2\Lambda_{ab}
\end{eqnarray}
with the matrices $G_{ab}$ and $\Lambda_{ab}$ solution of 
the Dyson equations
\begin{eqnarray}
\label{eq:dg2l}
  (G^{-1})_{ab} &=& \lambda\, \delta_{ab} - \beta^2Q_{ab}
                + G_{ab}\Lambda_{ab}
\\
\label{eq:dgam2l}
(\Lambda^{-1})_{ab} &=& \frac{1}{2}G_{ab}^2
\end{eqnarray}
and $G_{aa} = 1$.
Stationarity of ${\cal L}[Q]$ with respect to the variations of $Q_{ab}$ 
leads to the self-consistent equation $Q_{ab}=G_{ab}$, 
the form of which is dictated by the {\it ansatz} used for the matrix
$Q_{ab}$.

The Replicon component for the two-replica approximation reads:
\begin{eqnarray}
{\cal M}_{\hat{k}\hat{l}}^{rr}(p) &=& p^2 + 1
\nonumber\\
&\phantom{=}& - \beta^2\left[\frac{1}
{\widehat{Q}_k\widehat{Q}_l}
+\Lambda_r-Q_r\widehat{\Lambda}_k\widehat{\Lambda}_lQ_r\right]^{-1}
\end{eqnarray}
with $r = 0,\ldots, R$ and $k,l= r+1,\ldots, R+1$.

In the next subsection we examine the existence and stability of 
three possible scenarios, corresponding to three particular parameterizations
of the Parisi matrix: 
the Re\-pli\-ca Symmetric, the one step Replica Symmetry Breaking and
the full Replica Symmetry Breaking ansatz.

\subsection{The Replica Symmetric (RS) solution}
In the Replica Symmetric solution all replicas are treated on an equal footing,
and the matrices $Q_{ab}$ and $\Lambda_{ab}$ are given by
\begin{equation}
Q_{ab} = (1-Q)\,\delta_{ab} + Q, \quad
\Lambda_{ab} = (\Lambda_1 - \Lambda_0)\,\delta_{ab} + \Lambda_0 \ .
\end{equation}
Then from eq. (\ref{eq:dgam2l}) and the relation $Q_{ab} = G_{ab}$
we have for $n\to 0$
\begin{equation}
\Lambda_1-\Lambda_0 =  \frac{2}{1-Q^2}, \qquad
\Lambda_0 = -\frac{2\,Q^2}{(1-Q^2)^2}
\end{equation}
that, inserted back into (\ref{eq:dg2l}), leads to the RS equation
\begin{eqnarray}
\frac{1-2Q}{(1-Q)^2} &=& \lambda - \beta^2 + 2 \frac{1 -2Q^2}{(1-Q^2)^2}
\\
\frac{Q}{(1-Q)^2} &=& \beta^2Q + 2\frac{Q^3}{(1-Q^2)^2}\ .
\end{eqnarray}
These equations can be easily solved for $\lambda$ and $\beta$ as function of
$Q\in[0,1]$, and one ends up with
\begin{eqnarray}
\lambda &=& \frac{2\, Q}{1-Q^2} \\
\beta^2 &=& \frac{1+2Q - Q^2}{(1-Q^2)^2}\ .
\end{eqnarray}
By varying $Q$ between $0$ and $1$ one obtains the solution, if it exists, 
in the whole range of physical values of $Q$.
We note in particular that $Q$ vanishes as $T\to 1^-$, while 
$Q\to 1$ as $T$ goes to $0$:
\begin{equation}
Q = 1 - \frac{T}{\sqrt{2}} + O(T^2), \quad T\to 0\ .
\end{equation}
The stability of the RS solution is ruled by the Replicon eigenvalue
\begin{eqnarray}
{\cal M}^{00}_{\hat{1}\hat{1}}(p=0) &=& 1 - \frac{\beta^2}
        {(1-Q)^{-2} + \Lambda_0 - Q^2(\Lambda_1 - \Lambda_0)^2} 
\nonumber\\
         &=& \frac{4Q^2}{5 Q^2 - 2Q - 1}\ .
\end{eqnarray}
The Replicon eigenvalue is positive at $T=0$, where its value is $2$, and 
remains positive
up to the temperature $T^*\simeq 0.34...$, where the denominator vanishes.
For larger temperatures the Replicon is negative, and vanishes as $T\to 1^-$.
Therefore we conclude that the RS solutions exists for all temperatures
below $T=1$, as in the trivial case. But this solution is only stable 
for $T<T^*$. This is a rather unusual feature of this approximation.

\subsection{The one step Replica Symmetry Breaking (1RSB) solution}
In the previous section we have seen that the first correction coming from 
${\cal K}_2[G,\Lambda]$ is not enough to completely destabilize the 
RS solution that exists for $T<1$, yet its domain of validity is reduced to 
the temperature range up to $0.34\ldots$. In this section we investigate the 
existence of solutions parameterized by the so-called one step RSB ansatz:
\begin{equation}
Q_{ab} = (Q_2 - Q_1)\,\delta_{ab} + (Q_1 - Q_0)\,\epsilon_{ab} + Q_0
\end{equation}
where the matrix $\epsilon_{ab}$ is equal to $1$ if $a$ and $b$ are in anyone
of the diagonal boxes of size $p_1 \equiv m$, and $0$ otherwise. 
The diagonal element $Q_{R+1}=1$ of the overlap matrix writes $Q_2=1$.
The matrices $G_{ab}$ and $\Lambda_{ab}$ have a similar structure.

In the absence of an external field that breaks the up/down symmetry, 
$Q_0=0$.
This in turn implies that all $0$-indexed quantities, such as
$\Lambda_0$ and $G_0$, must also vanish. It is not difficult to show that 
$Q_0=\Lambda_0=G_0=0$ is a solution of the self-consistent equations of motion.

Inserting the 1RSB parameterization into eq. (\ref{eq:dgam2l}), using 
the condition $G_{ab} = Q_{ab}$,  and taking the $n\to 0$ limit, we have
\begin{eqnarray}
\Lambda_2-\Lambda_1 &=&  \frac{2}{1-Q_1^2}\ ,
\label{eq:1rsb1a}\\
\Lambda_1 &=& -\frac{2Q_1^2}{(1-Q_1^2)(1-Q_1^2 + m\,Q_1^2)}\ .
\label{eq:1rsb1b}
\end{eqnarray}
Note that for $m\to 0$ these relations reduce to those found for the RS case. 
This is not unexpected since if $Q_0=0$ the systems breaks down into 
$n/m$ subsystems of dimension $m$, each one having a RS structure, 
$Q_1$ playing the role of $Q$. For $m\to 0$ we recover the
RS solution.

The other Dyson equation, as of eq. (\ref{eq:dg2l}), leads to
\begin{eqnarray}
\lambda &=& \frac{(2-m)\, Q_1}{(1+Q_1)(1-Q_1 + m\,Q_1)} \\
\beta^2 &=& \frac{1 + 2 Q_1 + (m-1)\,Q_1^2}
               {(1+Q_1)(1+Q_1 -m\,Q_1)(1+Q_1^2 - m\,Q_1^2)}\ .
\label{eq:1rsb2}
\end{eqnarray}
As for the RS solution by varying $Q_1$ in the range $[0,1]$ we get,
for any value of $m\in [0,1]$ a 1RSB solution, if it exists.

To fix the value of $m$ we use the stationarity equation 
$(\partial/\partial m) {\cal L}[Q] = 0$ (yielding the so-called
{\sl static solution} \cite{KirThi87,CriSom92,CriHorSom93,Monasson95}). 
The computation is achieved 
using the formulae of Appendices \ref{app:RFT} and \ref{app:ldet}, leading to
\begin{eqnarray}
  \frac{\beta^2}{2}Q_1^2 
  &+& \frac{1}{m^2}\ln\left[\frac{1-Q_1}{1-Q_1+m\,Q_1}\right]
\nonumber\\
  &+& \frac{1}{m^2}\ln\left[\frac{\Lambda_2-\Lambda_1}
                     {\Lambda_2-\Lambda_1+m\,\Lambda_1}\right]
\nonumber\\
  &+& \frac{1}{m}\frac{Q_1}{1-Q_1+m\,Q_1} 
  + \frac{1}{m}\frac{\Lambda_1}{\Lambda_2-\Lambda_1+m\,\Lambda_1} 
\nonumber\\
  &-& \frac{1}{2}Q_1^2\,\Lambda_1 = 0\ .
\end{eqnarray}
By eliminating $\Lambda_1$, $\Lambda_2$ and $\beta$ in favor of $Q_1$ and 
$m$ with the help of eqs. (\ref{eq:1rsb1a})-(\ref{eq:1rsb1b}) and 
(\ref{eq:1rsb2}), we 
end up with the following equation 
\begin{eqnarray}
&\phantom{=}&
2 (1 - Q_1 + m\,Q_1)
\nonumber\\ &\phantom{=}& \hskip0.5cm 
\times (1+Q_1) \ln\left[
                    \frac{1-Q_1^2 + m\,Q_1^2}{(1+Q_1)(1-Q_1 + m\,Q_1)}
                              \right]
\nonumber\\
&\phantom{=}& \hskip0.5cm
+ m\,Q_1(2 + m\,Q_1) = 0
\end{eqnarray}
that gives $m\equiv m(Q_1)$ for the 1RSB case.
Solving this equation shows that $m(Q_1)$ is a monotonically increasing 
function of $Q_1$ that vanishes for $Q_1=0$, i.e., $T=1$, while 
$\lim_{Q_1\to 1} m(Q_1) = 4\ln 2 - 2 < 1$.

It turns out that the 1RSB solution appears at the critical temperature $T=1$
and exists down to the lower critical temperature 
$T_{\rm 1RSB} \simeq 0.656485\ldots$ where $Q_1$ reaches the maximum allowed
value $Q_1 = 1$. 

The stability of the 1RSB solution is now ruled by two Replicon eigenvalues,
\begin{equation}
{\cal M}^{00}_{\hat{1}\hat{1}}(p=0)  = 1 - \beta^2(1-Q_1)^{2} 
\end{equation}
and
\begin{eqnarray}
{\cal M}^{11}_{\hat{2}\hat{2}}(p=0)  &=& 1 \\
&\phantom{=}& - \frac{\beta^2}
         {(1-Q_1)^{-2} + \Lambda_1 - Q_1^2(\Lambda_2 -\Lambda_1)^{-2} }\ .
\nonumber
\end{eqnarray}
At $T=1$ both eigenvalues vanish. These two eigenvalues cross each other
at $T_0 = 0.6760$. Above $T_0$, ${\cal M}^{11}_{\hat{2}\hat{2}}$ is the lowest
eigenvalues and is negative. Below $T_0$, ${\cal M}^{00}_{\hat{1}\hat{1}}$ is 
also negative. Therefore the solution is always unstable.

\subsection{The full Replica Symmetry Breaking (full-RSB) solution}
We consider now the solution, if any, with an
infinite number of replica symmetry breaking steps (full-RSB). 
In this limit the matrices $Q_{ab}$, $G_{ab}$ and $\Lambda_{ab}$ 
are described by functions of a single parameter $x$ varying between 
$0$ and $1$. 
To find the self-consistent equation of motion of the full-RSB solution let us
consider the case of $R$ replica symmetry breaking steps. The full-RSB 
solution is obtained as the limit $R\to\infty$.

By assuming a Parisi's $R$ replica symmetry breaking structure for the matrices 
$Q_{ab}$, $G_{ab}$ and $\Lambda_{ab}$, the Dyson equations (\ref{eq:dg2l}) and
(\ref{eq:dgam2l}) for the non-diagonal terms becomes
\begin{eqnarray}
\label{eq:dyriq}
(G^{-1})_r &=& -\beta^2\,Q_r + Q_r\,\Lambda_r\ , \\
\label{eq:dyrig}
(\Lambda^{-1})_r &=& \frac{1}{2}Q_r^2\ , \qquad\qquad\qquad r = 0, \ldots, R\ .
\end{eqnarray}
These equations are solved through the Replica Fourier Transform (RFT) 
(see Appendix \ref{app:RFT} for proper definition and Appendix \ref{app:ldet}
for an example). 
The RFT of eq. (\ref{eq:dyrig}) reads
\begin{equation}
\label{eq:lhm1}
\widehat{\Lambda^{-1}}_k = \frac{1}{\widehat{\Lambda}_k} = 
           \frac{1}{2}\widehat{q}_k
\end{equation}
where we have introduced the shorthand $q_{ab} = (Q_{ab})^2$.
Inverting the RFT yields in the continuous limit $R\to\infty$ 
\begin{eqnarray}
\Lambda(x) &=& -\int_{0}^{x} ds\, \frac{1}{s}\frac{d}{ds}\widehat{\Lambda}(s)
\nonumber\\
           &=& 2\int_{0}^{x} ds\, \frac{1}{\widehat{q}(s)^2}
                               \frac{d}{ds}\widehat{q}(s)
\label{eq:lx}
\end{eqnarray}
where,
\begin{equation}
\label{eq:q2h}
\widehat{q}(x) = 1 - Q(x_c)^2 + \int_{x}^{x_c} ds\,s\,\frac{d}{ds}Q(s)^2
\end{equation}
and
\begin{equation}
x_c = \lim_{R\to\infty} p_R, \quad Q(x_c) = \lim_{R\to\infty} Q_R\, .
\end{equation}
Finally, from eqs. (\ref{eq:dyriq}), (\ref{eq:lx}) and (\ref{eq:q2h}) we have
the equation of motion for $Q(x)$ for $0\leq x\leq x_c$:
\begin{equation}
\label{eq:Q2l}
\int_{0}^{x}ds\, \frac{\dot{Q}(s)}{\widehat{Q}(s)^2} =
                    \beta^2\,Q(x) + 4\, Q(x)
     \int_{0}^{x}ds\, \frac{Q(s)\,\dot{Q}(s)}{\widehat{q}(s)^2}
\end{equation}
where the ``dot'' denotes the derivative of the function with respect to
its argument.
As done for the 1RSB solution we assumed that all $0$-indexed quantities,
such as $Q_0\equiv Q(0)$ etc., vanish.

The solution of this integral equation is obtained by successive 
differentiation with respect to $x$.
The first yields
\begin{equation}
\label{eq:dQ2l}
\frac{1}{\widehat{Q}(x)^2} =
                    \beta^2 + 4
     \int_{0}^{x}ds\, \frac{Q(s)\,\dot{Q}(s)}{\widehat{q}(s)^2}
        + 4 \frac{Q(x)^2}{\widehat{q}(s)^2}\ .
\end{equation}
By setting $x=0$ we get
\begin{equation}
\label{eq:rel1}
\widehat{Q}(0) = 1 - Q(x_c) + \int_{0}^{x_c}ds\, s\,\dot{Q}(s) = T\ .
\end{equation}
As $Q(x) = Q(x_c)$ for $x\geq x_c$, 
equation (\ref{eq:rel1}) reads
\begin{equation}
\label{eq:sumr1}
   1 - \int_{0}^{1}dx\, Q(x) = T\ .
\end{equation}
Notice that 
the same manipulations applied to the internal energy (per site) $u$, yield
\begin{equation}
\label{eq:sumr2}
\frac{1}{2}\left(1 - \int_{0}^{1}dx\, Q^2(x)\right) = -u T\ .
\end{equation}
The sum rules (\ref{eq:sumr1}) and (\ref{eq:sumr2}) are known to be exact for
the SK model. They are verified here to the first order of the 2-PI expansion
of ${\cal K}_2$.

\begin{figure}
\includegraphics[scale=0.6]{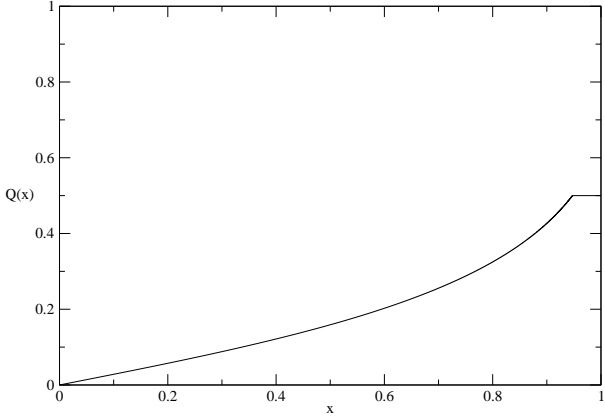}
\caption{$Q(x)$ versus $x$ for the full-RSB solution.
         The solution is obtained with $Q_c = 0.5$ and corresponds
         to temperature $T=0.8076\ldots$}
\label{fig:2rpl-qx05}
\end{figure}

To solve the equation of motion (\ref{eq:dQ2l}) we take one more 
$x$ derivative to get
\begin{equation}
\label{eq:xq}
x = 6\,Q(x)\, \frac{\widehat{Q}(x)^3\,\widehat{q}(x)}
            {\widehat{q}(x)^3 - 8\,Q(x)^3 \widehat{Q}(x)^3}\ .
\end{equation}
Taking advantage of 
\begin{eqnarray}
\widehat{Q}(Q) &=& 1 - Q_c + \int_{Q}^{Q_c}dQ'\, x(Q')
\nonumber\\
\label{eq:qhq}
               &=& 1 - Q_c - v(Q)
\end{eqnarray}
\begin{eqnarray}
\widehat{q}(Q) &=& 1 - Q_c^2 + 2 \int_{Q}^{Q_c}dQ'\, Q'\, x(Q')
\nonumber\\
\label{eq:q2hq}
               &=& 1 - Q_c^2 -2 Q\,v(Q) + 2\,r(Q)
\end{eqnarray}
where $Q_c = Q(x_c)$ is the {\it plateau} value, 
it is easy to check that eqs. (\ref{eq:xq}), (\ref{eq:qhq}) and 
(\ref{eq:q2hq}) are equivalent to the coupled differential equations
\begin{eqnarray}
\label{eq:sis1}
\frac{d}{dQ}r(Q) = v(Q) \nonumber\\
\frac{d}{dQ}v(Q) = x(Q)
\end{eqnarray}
with the boundary condition
\begin{equation}
r(Q_c) = v(Q_c) = 0.
\end{equation}
Solving these equations for $0\leq Q_c \leq 1$, and
fixing the temperature through eq. (\ref{eq:rel1}), we obtain the complete 
solution $Q(x)$.
In Fig. \ref{fig:2rpl-qx05} we show the form of $Q(x)$.
As found for the 1RSB solution, the temperature does not vanishes 
as $Q_c\to 1$, but reaches the finite value 
$\lim_{Q_c\to1}T(Q_c) = T_{\rm full-RSB} = 0.7829\ldots$.

At variance with the 1RSB solution the continuous solution
is marginally stable since the lowest Replicon eigenvalues vanishes.
Indeed, looking at stability, we have for the Replicon eigenvalues:
\begin{eqnarray}
\label{eq:cri}
{\cal M}_{\hat{k}\hat{l}}^{x x}(p) &=& p^2 + 1
\\
&\phantom{=}& - \beta^2\left[
  \frac{1}{\widehat{Q}_k\widehat{Q}_l}
  + \left(\frac{2}{q}\right)_x 
   - 4Q_x^2\frac{1}{\widehat{q}_k\widehat{q}_l}\right]^{-1}\ .
\nonumber
\end{eqnarray}
Before letting $R\to\infty$ the lowest eigenvalue is 
${\cal M}^{rr}_{\widehat{r+1}\widehat{r+1}}$. In the continuum  
${\cal M}^{xx}_{\widehat{x+}\widehat{x+}}$ is correspondingly given by,
\begin{eqnarray}
\label{eq:cry}
{\cal M}_{\widehat{x+}\widehat{x+}}^{x x}(p) &=& p^2 + 1
\\
&\phantom{=}& \hskip-0.5cm
-\beta^2\left[
        \left(\frac{1}{\widehat{Q}_x}\right)^2 + 
         \left(\frac{2}{q}\right)_x
        - 4Q_x^2\left(\frac{1}{\widehat{q}_x}\right)^2
          \right]^{-1}\ .
\nonumber
\end{eqnarray}
On (\ref{eq:cry}), one identifies, from the equation of motion 
(\ref{eq:dQ2l}), that 
the quantity in the square bracket is equal to $T^2$, leading to a
zero mode:
\begin{eqnarray}
{\cal M}_{\widehat{x+}\widehat{x+}}^{x x}(p) = p^2\ .
\end{eqnarray}
This is a check of the existence of Goldstone zero modes, 
accompanying the breakdown of a continuous invariance group, {\it i.e.} the 
existence of Ward-Takahashi identities \cite{DeDomTemKon98}. 
From (\ref{eq:cri}) one can also 
obtain the size of the band on top of the zero-modes 
${\cal M}_{\widehat{x+}\widehat{x+}}^{x x}$ by evaluating:
\begin{eqnarray}
{\cal M}_{\widehat{x_c}\widehat{x_c}}^{00}
  -{\cal M}_{\widehat{0+}\widehat{0+}}^{0 0}&=&
    {\cal M}_{\widehat{x_c}\widehat{x_c}}^{x x} \nonumber\\
    &=&
    1-\frac{1}{T^2}\frac{1}{\big(\frac{1}{1-Q_1}\big)^2}
\nonumber\\
       &=& 1-\beta^2(1-Q_1)^2\ .
\end{eqnarray}
As a result at this level of approximation, one keeps the same structure 
as for the SK model near $T_c$: a Replicon broad band of modes, bordered below
by zero modes.

\begin{figure}
\includegraphics[scale=0.8]{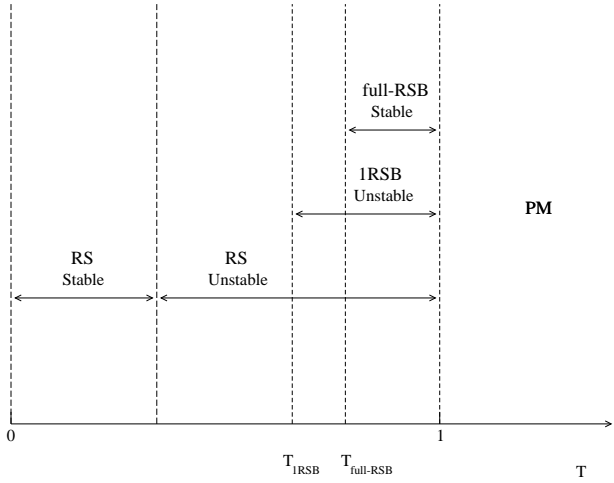}
\caption{Phase diagram in the two-replica approximation.
         The RS solution ($R=0$) exists for all temperature below $T=1$,
         but it is stable below $T^*=0.34\ldots$ only.
         The 1RSB solution ($R=1$) exists in the temperature range
         $T_{\rm 1RSB} = 0.656\ldots \leq T <1$, 
           but it is unstable. Finally the full-RSB solution ($R\to\infty$)
         exists in the temperature range $T_{\rm full-RSB}\leq T<1$, 
         $T_{\rm full-RSB} = 0.783\ldots$, and it is marginally stable. 
         Above $T=1$ only the paramagnetic (PM) solution with 
         $Q_{ab}=\delta_{ab}$ exists. 
         All solutions coincide at $T=1$.}
\label{fig:2rpl-ph-diag}
\end{figure}

\section{Conclusion}
In this work, we have introduced, for disordered Ising-like systems, an 
expansion around their spherical approximation in replica space via the use of 
constraint-fields and 
a double Legendre Transform of the free energy functional. 
Successive approximations are given by 2-PI graphs constructed with replicated
spin correlators and replicated constraints correlators.
A striking feature of this approach is that one can keep track of the the
matrix structure of the order parameter (and thus of the Hessian) at all
temperatures.

We have applied the method to the SK model 
and analyzed in detail the lowest nontrivial contribution, studying in each
case (RS, 1RSB, full-RSB) successively, the existence of solutions and 
their stability as summarized in Fig. \ref{fig:2rpl-ph-diag}. At high enough
$T$, the constraint-field fluctuations are sufficient to destabilize the
RS solution giving rise, instead, to a marginally stable continuous solution.
At low enough $T$ the constraint-field becomes ineffective: a stable RS
solution is found.

The ineffectiveness of the constraint-field seems linked to the fact that,
at its lowest order the 2-PI functional is linear in the constraint-field.
This leads actually to a trivial relationship between the constraint-field 
and the order 
parameter and a correction to the free energy that is linear in temperature, a 
very peculiar feature of the lowest order approximation.
This remark motivates
future work to take into account more generic features of the 
model as in the four replica approximation.

\begin{acknowledgement}
One of us (C.D.) would like to thank E. Brezin and H. Orland for useful 
discussions. A.C. would like to thank the IPhT of CEA, 
where part of this work was done, for hospitality and support.
\end{acknowledgement}

\appendix
\section{Inversion of a Parisi Matrix with the RFT formalism}
\label{app:RFT}
In the $R$ steps RSB scheme an $n\times n$ 
symmetric matrix $Q_{ab}$ is parameterized by two sets of numbers. The first 
set $(p_0,p_1,\ldots,p_R, p_{R+1})$, with $p_0=n$ and $p_{R+1}=1$, 
gives the size of the boxes the matrix is successively divided into. 
By definition for any finite $n$ the $p_r$  are decreasing integer numbers
$p_r > p_{r+1}$, however in the $n\to 0$ limit they becomes real numbers in
the range $[0,1]$ and its order gets reversed, that is, $p_{r+1} > p_r$.
The second set
$(Q_0, Q_1,\ldots,Q_R, Q_{R+1})$, gives the value of $Q_{ab}$ in each box, 
\begin{equation}
Q_{ab} \equiv Q_{a\cap b=r} = Q_r, \qquad r = 0, \cdots, R+1\ .
\end{equation}
The notation $a\cap b=r$ means that the indexes $a$ and $b$ belong to the same
box of size $p_r$ but to two {\sl distinct} boxes of size $p_{r+1}$. 
The term $Q_{R+1}$ is the value of diagonal term $Q_{aa}$.

The Replica Fourier Transform $\widehat{Q}_{ab}$ of the matrix $Q_{ab}$ is 
defined by \cite{DeDomCarTem97,note:rft}:
\begin{equation}
\widehat{Q}_k = \sum_{r=k}^{R+1}\,p_r\,(Q_r - Q_{r-1}), 
\qquad k = 0,1,\ldots, R+1
\end{equation}
where we have taken $Q_{-1}=0$. In the following all quantities with index out 
of the allowed range are assumed to be equal to zero.
From the above definition, the useful relation
\begin{equation}
\label{eq:arel}
\widehat{Q}_k - \widehat{Q}_{k+1} = p_k\,(Q_k - Q_{k-1})
\end{equation}
allows the inversion of the RFT
\begin{equation}
\label{eq:arfti}
Q_r = \sum_{k=0}^{r}\,\frac{1}{p_k}\,(\widehat{Q}_k - \widehat{Q}_{k+1}), 
\quad r = 0,1,\ldots, R+1\ .
\end{equation}
It can be shown \cite{DeDomCarTem97} that the Replica Fourier Transform has the
properties of the standard Fourier Transform. In particular the RFT
of the inverse matrix $(Q^{-1})_{ab}$ is the inverse of the RFT of $Q_{ab}$:
\begin{equation}
\widehat{Q^{-1}}_k = \frac{1}{\widehat{Q}_k}\ .
\end{equation}
Then, by virtue of eqs. (\ref{eq:arel}) and (\ref{eq:arfti}) we have
for $r = 0,\cdots, R$,
\begin{eqnarray}
(Q^{-1})_r &=& \sum_{k=0}^{r} \frac{1}{p_k}\,\left(
           \frac{1}{\widehat{Q}_k} - \frac{1}{\widehat{Q}_{k+1}}
             \right)
\nonumber\\
           &=& - \sum_{k=0}^{r} \,
              \frac{Q_k - Q_{k-1}}{\widehat{Q}_{k+1}\,\widehat{Q}_k}
\end{eqnarray} 
and
\begin{eqnarray}
(Q^{-1})_{R+1} &=& \frac{1}{\widehat{Q}_{R+1}} + (Q^{-1})_{R}
\nonumber\\
               &=& \frac{1}{Q_{R+1}-Q_R} 
                 - \sum_{k=0}^{R} \,
              \frac{Q_k - Q_{k-1}}{\widehat{Q}_{k+1}\,\widehat{Q}_k}\ .
\end{eqnarray}
These expression remains valid also in the $n\to 0$ limit.

In the limit of $R\to\infty$ the difference $Q_r - Q_{r-1}$ with 
$r \leq R$ goes to zero,  
the sum can then be replaced by an integral and we have,
for $0\leq x\leq x_c\equiv p_R$,
\begin{equation}
(Q^{-1})(x) = -\int_{0}^{x}\, ds \frac{\dot{Q}(s)}{\widehat{Q}(s)^2} 
               - \frac{Q(0)}{\widehat{Q}(0)^2}, 
\end{equation}
where 
\begin{equation}
\widehat{Q}(x) = Q(1) - Q(x_c) + \int_x^{x_c}\,ds\, s\,\dot{Q}(s)\ .
\end{equation}
and $Q(1) \equiv Q_{R+1}$.
Similarly
\begin{eqnarray}
  (Q^{-1})(1) &=& 
    \frac{1}{Q(1) - Q(x_c)}  + (Q^{-1})(x_c) \\
&=& 
    \frac{1}{Q(1) - Q(x_c)}  
    -\int_{0}^{x_c}\, ds \frac{\dot{Q}(s)}{\widehat{Q}(s)^2} 
               - \frac{Q(0)}{\widehat{Q}(0)^2}\ .
\nonumber
\end{eqnarray}

\section{Computing ${\rm Tr\ln Q}$}
\label{app:ldet}
The quantity ${\rm Tr\ln Q}$ can be computed by using the identity
\begin{equation}
\frac{1}{n}{\rm Tr}\ln Q = \lim_{m\to 0}\frac{1}{m}
                   \left(\frac{1}{n}{\rm Tr}Q^m - 1\right)
\end{equation}
and the results of the previous Appendix.

The quantity $Q^m$ for integer $m$ is the matrix product of $m$ matrices, i.e.,
a convolution in matrix indexes. As a consequence
\begin{equation}
\widehat{Q^m}_r = \widehat{Q}_r^m\ .
\end{equation}
Then, by using eq. (\ref{eq:arfti}), we have
\begin{equation}
\frac{1}{n}{\rm Tr}Q^m = (Q^m)_{R+1} = \sum_{k=0}^{R+1}\frac{1}{p_k}
                     \left(\widehat{Q}_k^m - \widehat{Q}_{k+1}^m\right)
\end{equation}
and hence
\begin{eqnarray}
\frac{1}{n}{\rm Tr}\ln Q &=& \sum_{k=0}^{R+1}\frac{1}{p_k}
                     \left(\ln\widehat{Q}_k - \ln\widehat{Q}_{k+1}\right)
\nonumber\\
&\phantom{=}& \hskip-0.5cm
                 \mathop{=}_{n\to 0}
                    \ln(Q_{R+1} - Q_R) 
             + \sum_{k=1}^R\frac{1}{p_k}\,\ln\left(
                   \frac{\widehat{Q}_k}{\widehat{Q}_{k+1}}
                                           \right)
\nonumber\\
&\phantom{=}& 
             + \frac{Q_0}{\widehat{Q}_1}\ .
\label{eq:atrlg}
\end{eqnarray}
This expression is valid for any finite $R$. To perform the $R\to\infty$ limit
we use the relation (\ref{eq:arel}) and expand the logarithm in the sum for
small $Q_r-Q_{r-1}$, i.e.,
\begin{eqnarray}
\ln\left(\frac{\widehat{Q}_k}{\widehat{Q}_{k+1}}\right) &=&
\ln\left(1 + p_k\frac{Q_k - Q_{k-1}}{\widehat{Q}_{k+1}}\right) 
\\
&=& 
         p_k\frac{Q_k - Q_{k-1}}{\widehat{Q}_{k}} 
         + O\bigl((Q_{k} - Q_{k+1})^2\bigr)
\nonumber
\end{eqnarray}
so that eq. (\ref{eq:atrlg}) becomes
\begin{eqnarray}
\frac{1}{n}{\rm Tr}\ln Q &=& \ln(Q_{R+1} - Q_R)
             + \sum_{k=1}^R
                     \frac{Q_k - Q_{k-1}}{\widehat{Q}_{k}}             
\nonumber\\
&\phantom{=}&
             + \frac{Q_0}{\widehat{Q}_1}
             + \sum_{k=1}^R\,O\bigl((Q_{k} - Q_{k+1})^2\bigr)\ .
\end{eqnarray}
In the limit $R\to\infty$ the last sum vanishes and we end up with
\begin{eqnarray}
\frac{1}{n}{\rm Tr}\ln Q &=& \ln\bigl(Q(1) - Q(x_c)\bigr) 
\nonumber\\
&\phantom{=}& 
             + \int_{0}^{x_c}\,dx\,
                     \frac{\dot{Q}(x)}{\widehat{Q}(x)}             
             + \frac{Q(0)}{\widehat{Q}(0)}
\end{eqnarray}

\section{Proof of eq. (\ref{eq:RSsol})}
\label{app:proof}
If we neglect the term ${\cal K}_2[G,\Lambda]$, and use
$G_{ab} = Q_{ab}$ the functional ${\cal L}$ can be written as
\cite{CriSom92,CriLeu06}:
\begin{eqnarray}
\frac{2}{n} {\cal L} &=& \frac{\beta^2}{2n}\sum_{ab}Q_{ab}^2 
            +\frac{1}{n} {\rm Tr}\ln Q + {\rm const}
\nonumber\\
           &=& \beta^2\int_{0}^{1}dQ\, Q\,x(Q) 
             + \int_{0}^{Q_R}\frac{dQ}{\int_{Q}^1dQ' x(Q')} 
\nonumber\\
&\phantom{=}&
             + \ln (1 - Q_R) + {\rm const}
\end{eqnarray}
where
\begin{equation}
  x(Q) = p_0 + \sum_{r=0}^R (p_{r+1}-p_r)\,\theta(Q-Q_r)\ .
\end{equation}
The saddle point equations are obtained by varying
the above functional with respect to $x(Q)$:
\begin{equation}
\frac{2}{n}\,{\cal L} = \int_{0}^{1}\,dQ\, F(Q)\,\delta x(Q)
\end{equation}
where
\begin{equation}
\label{eq:sp}
F(Q) = \beta^2\, Q - \int_{0}^{Q} \frac{dQ'}
                  {\left[\int_{Q'}^{1} dQ''\, x(Q'')\right]^2}
\end{equation}
and
\begin{eqnarray}
 \delta x(Q) &=& \sum_{r=0}^{R}\,(\delta p_{r+1} - \delta p_r)\,
                     \theta(Q - Q_r)
\nonumber \\
&\phantom{=}& 
    - \sum_{r=0}^{R}\,(p_{r+1} - p_r)\,\delta(Q-Q_r)\,\delta Q_r.
\end{eqnarray}
By requiring the stationarity of ${\cal L}$ with respect to variations of
$Q_r$ and $p_r$ one gets, respectively
\begin{equation}
\label{eq:fq0}
F(Q_r) = 0, \qquad r = 0,\ldots, R\ , 
\end{equation}
\begin{equation}
\label{eq:mst}
\int_{Q_{r-1}}^{Q_r}\, dq\, F(Q) = 0, \qquad r = 1,\ldots, R\ .
\end{equation}
The function $F(Q)$ is continuous, thus Eq. (\ref{eq:mst}) implies
that between any two successive $Q_r$ there must be at least two
extrema of $F(Q)$.  If we denote these by $Q^*$, then the extremal
condition $F'(Q^*)=0$ implies that
\begin{equation}
\label{eq:der}
\int_{Q^*}^{1}\, dq\, x(q) = T\ .
\end{equation}
The right hand side of this equation is a constant and hence there is 
only one solution (or none). Thus the only solution of 
eq. (\ref{eq:mst}) is $Q_{r-1} = Q_r = Q$ for all $r$. This in turn implies 
that $x(Q') = \theta(Q'-Q)$, so that eq. (\ref{eq:fq0}) reduces to 
\begin{equation}
\beta^2\, Q - \frac{Q}{(1-Q)^2} = 0
\end{equation}
that is, $Q=1-T$.

\section{Double Legendre Transform}
\label{app:DLT}
In this section we give a short summary of the double Legendre Transform
to help to understand the origin of eq. (\ref{eq:l}). To keep the 
notation as simple as possible we shall consider the case of a single scalar
field. The extension to more complicated cases is straightforward.

We are interested in the evaluation of
\begin{equation}
\ln Z = \ln\int\, d\phi\, e^{-S[\phi]}
\end{equation}
for some $S[\phi]$. To this end we define the generating function
$W[J,K]$ as \cite{CorJacTom74,Haymaker91}
\begin{equation}
e^{W[J,K]} =
\int\, d\phi\, e^{-S[\phi] + J\phi + \frac{1}{2}\phi K\phi}\ .
\end{equation}
Clearly
\begin{equation}
\label{eq:lnZ}
  \ln Z =  W[J,K]\Bigr|_{J=K=0}\ .
\end{equation}
From the definition of $W[J,K]$ it follows
\begin{eqnarray}
\frac{\partial}{\partial J} W[J,K] &=& \langle\phi\rangle = \varphi
\\
\frac{\partial}{\partial K} W[J,K] &=& \frac{1}{2}\langle\phi\phi\rangle 
                         = \frac{1}{2}\left(
         \varphi\,\varphi + G
            \right)
\end{eqnarray}
where the average $\langle\cdots\rangle$ is taken with the full weight
\begin{equation}
\exp\left[-S[\phi] + J\phi +\frac{1}{2}\phi K \phi\right]
\end{equation}
 and $G$ is the connected
two-point correlator. We can then define the double Legendre Transform 
$\Gamma[\varphi,G]$ of $W[J,K]$ as
\begin{equation}
\label{eq:dlt1}
W[J,K] + \Gamma[\varphi,G] = J\varphi + \frac{1}{2}\varphi K\varphi 
                              + \frac{1}{2} K G
\end{equation}
with
\begin{eqnarray}
\label{eq:dlt2}
\frac{\partial}{\partial \varphi} \Gamma[\varphi,G] &=& J + K\varphi\ ,
\\
\label{eq:dlt3}
\frac{\partial}{\partial G} \Gamma[\varphi,G] &=& \frac{1}{2} K \ .
\end{eqnarray}
Comparison of eq. (\ref{eq:lnZ}) with eqs. (\ref{eq:dlt1})-(\ref{eq:dlt3})
shows that
\begin{equation}
\ln Z = -\Gamma[\varphi,G]
\end{equation}
where $\varphi$ and $G$ are solution of the equations
\begin{eqnarray}
\frac{\partial}{\partial \varphi} \Gamma[\varphi,G] &=& 0
\\ 
\frac{\partial}{\partial G} \Gamma[\varphi,G] &=& 0
\end{eqnarray}
i.e., the value of $\ln Z$ is equal to (minus) the value of 
$\Gamma[\varphi,G]$ at its stationarity point.

Up to now we have just used some general properties of the Legendre
Transform. The usefulness of this approach arises from the fact that
one has an explicit form of $\Gamma[\varphi,G]$. Indeed one has
\begin{eqnarray}
\Gamma[\varphi,G] &=& S[\varphi] + \frac{1}{2}{\rm Tr}\ln G^{-1} 
                   + \frac{1}{2}{\rm Tr}D^{-1}G 
\nonumber\\
  &\phantom{=}& + {\cal K}_2[\varphi, G]
                   -\frac{1}{2}{\rm Tr} 1
\end{eqnarray}
where
\begin{equation}
D^{-1}[\varphi] = \left.\frac{\partial^2}{\partial\phi\,\partial\phi}S[\phi]
         \right|_{\phi=\varphi}
\end{equation}
and ${\cal K}_2[\varphi,G]$ is given by the sum of all 2-PI vacuum diagrams
of a theory with propagators $G$ and interaction vertices 
determined by the potential $V_{\rm int}$ given by
\begin{equation}
    S[\varphi+\phi] - S[\varphi] 
    - \phi\,\left.\frac{\partial}{\partial\phi}S[\phi]\right|_{\phi=\varphi} =
\frac{1}{2}\phi D^{-1}[\varphi]\phi + V_{\rm int}[\phi;\varphi]\ .
\end{equation}
This procedure corresponds to a dressed loop expansion with vertices that depend
on $\varphi$ and can thus exhibit non-perturbative effects even for a small
number of dressed loops. The stationarity condition on $\Gamma[\varphi,G]$ 
yields a set of coupled (nonlinear) equations for $\varphi$ and $G$.


\begin{thebibliography}{99}
\bibitem{note:omegaq}
  Notice that $\Omega[Q]$ is disorder independent. If we were to study another 
  $2$-spin model (ROM, etc\ldots), the Lagrangean would read 
  ${\cal L}[Q] = G[Q] + \Omega[Q]$ where $\Omega[Q]$ is as defined in 
  eq. (\protect\ref{eq:ome}) and $G[Q]$ determined as in Ref. 
  \protect\cite{refY}. The 
  Lagrangean can be extended to more complex spin interactions following
  Ref. \protect\cite{CriLeuPar02}. In this work we keep to the SK model.
\bibitem{note:slcorr}
  If one keeps the $\langle\sigma_a\lambda_b\rangle$ 
  correlators then their equation of motion only admits the trivial null
  solution.
\bibitem{note:lambda}
 We leave aside the possibility of having vectorial replica symmetry breaking.
\bibitem{note:q0}
  For instance in the $q\geq 4$ Potts model the symmetry is broken and indeed
  $Q_0\not=0$.
\bibitem{note:rft}
  The RFT was first introduced, directly in the continuum limit 
  ($R\to\infty$) by Mezard and Parisi\protect\cite{MezPar91}
\end{thebibliography}
\end{document}